\documentclass[manuscript]{aastex}

\usepackage{natbib}
\newcommand{\hh}{\textrm{H}_2}

\newcommand{\nhtot}{n_{\textrm{H}}}
\newcommand{\nhh}{n(\textrm{H}_2)}
\newcommand{\nh}{n(\textrm{H})}

\newcommand{\colhh}{N(\textrm{H}_2)}
\newcommand{\colh}{N(\textrm{H})}
\newcommand{\hcol}{N_\textrm{H}}

\newcommand{\fhh}{f(H_2)}
\newcommand{\rh}{R_{H_2}}

\newcommand{\lya}{\textrm{Ly}\alpha}

\slugcomment{To be submitted to ApJ}

\shorttitle{Propagation of Ly-$\alpha$ in protoplanetary disks}
\shortauthors{Bethell  et al}

\begin{document}

\title{The propagation of Ly-$\alpha$ in evolving protoplanetary disks}

\author{Thomas J. Bethell\altaffilmark{1} \& Edwin A. Bergin\altaffilmark{1}}
\affil{Department of Astronomy, University of Michigan, 500 Church St.,Ann Arbor, MI, 48109, USA }
\email{tbethell@umich.edu}

\begin{abstract}
We study the role resonant scattering plays in the transport of $\lya$ photons in accreting protoplanetary disk systems subject to varying degrees of dust settling. While the intrinsic stellar FUV spectrum of accreting
T Tauri systems may already be dominated by a strong, broad $\lya$ line ($\sim$80\%
of the FUV luminosity), we find that resonant scattering further enhances the $\lya$ density in the deep molecular layers of the disk.  $\lya$ is scattered downwards efficiently by the photodissociated atomic hydrogen layer that exists above the molecular disk.  In contrast, FUV-continuum photons pass unimpeded through the photodissociation layer, and (forward-)scatter inefficiently
off dust grains.  Using detailed, adaptive grid Monte Carlo radiative transfer simulations we show that the resulting $\lya$/FUV-continuum photon density ratio is strongly stratified; FUV-continuum dominated in
the photodissociation layer, and $\lya$-dominated field in the molecular disk. The enhancement is greatest
in the interior of the disk ($r\sim1$AU) but is also observed in
the outer disk ($r\sim100$AU). The majority of the total disk mass
is shown to be increasingly $\lya$ dominated as dust settles towards
the midplane.
\end{abstract}


\section{Introduction\label{sec:Introduction}}

T Tauri stars (TTS) are pre-main sequence low mass stars frequently
described as young analogs of our Solar System. Surrounded by 
circumstellar disks of gas and dust, these systems mark the earliest
stages of planet-formation, a process now understood to be
common in our Galaxy \citep{Adams:1987qf,Kenyon:1995uq,Pollack:1996fk}. Due to their close proximity to the parent star,
intense ultraviolet and X-ray radiation fields greatly impact the evolution of the protoplanetary disk.  The far-ultraviolet (FUV, $h\nu<13.6$eV) field is of particularly broad interest as it provides thermodynamical gas heating through the photoelectric effect \citep{Weingartner:2001jl}, and affects disk chemistry directly through the photo-dissociation and -ionization of molecules and atoms \citep[][and references therein]{Bergin:2007dz}. FUV photodesorption of ices contributes to the transport of material between solid and gas phases, whilst enabling complex molecule formation on grains surfaces \citep{Oberg:2009kh,Oberg:2009cq}.  Alongside X-rays and cosmic-rays, the FUV field is also involved in the ionization balance of the disk; of central importance in dynamical processes such as the magneto-rotational instability that is believed to play a role in accretion \citep{Balbus:1991fv,Gammie:1996ff}. Eventually the dispersal of disk gas may be driven by UV photo-evaporation \citep{Alexander:2006bs}, although resolving the contributions made by the separate FUV, EUV and X-ray components is still a matter of debate \citep{Gorti:2009ve,Owen:2010ly}


One feature of the FUV spectra in T Tauri systems which separates the accretors from the non-accretors,  is a strong, broad $\lya$ line (accounting for as much as 80\% of the total FUV luminosity). In nearby systems such as TW Hydra the $\lya$ line can be detected directly
\citep{Herczeg:2002fu}, whilst in obscured systems its presence may be inferred through the excitation of $\hh$ molecules \citep{Herczeg:2004kb,Bergin:2004if}. It was soon realized that the concentration of UV photons in a relatively narrow wavelength range ($1215.67\pm1$ \AA) would have important implications for wavelength-dependent chemical processes, specifically the photodissociation of molecules \citep{Bergin:2003la}.  For example, the molecules HCN, OH, H$_2$O and H$_2$CO can all be photodissociated by a $\lya$ photon, unlike the closely related species CN, OH$^+$, H$_2$O$^+$ and HCO$^+$ \citep[for a more complete list, see ][]{vDishoeck:2006kc}.  Abundance ratios frequently used as chemical diagnostics, such as $N(\textrm{CN})/N(\textrm{HCN})$,  will be changed (in this case increased) by $\lya$-dominated photodissociation.  \citet{Fogel:2011ys} have shown quantitatively that the inclusion of differential photodissociation by $\lya$ can  some instances help bring abundance ratios into better agreement with observation \citep{Thi:2004vn}. While the the presence of $\lya$ decreases the abundance of  HCN, NH$_3$, and CH$_4$ by an order of magnitude or more, other species such as CO$_2$ and SO are enhanced.  SO actually has a significant cross-section at 1216$\AA$, however so does SO$_2$, and it is the photodissociation of SO$_2$ which partially replenishes the SO abundance.  In this instance, the presence of $\lya$ actually increases the SO abundance.  Similarly, the abundances of H$_2$O and OH in the \citet{Fogel:2011ys} models increased with the introduction of $\lya$, even though both H$_2$O and OH have a significant photodissociation cross-sections at 1216$\AA$.  In this case the gas phase abundances are more than adequately replenished by efficient $\lya$ photodesorption from grain surfaces.  Further chemical modeling is required before the implications of $\lya$-dominated photochemistry are fully understood.

Despite the interesting chemical implications of a $\lya$ dominated FUV field, the problem of precisely how it penetrates the disk has received little attention in the literature.  In this paper we attempt to shed light upon its basic transport.  Although frequently flared, protoplanetary disks remain relatively flat objects with vertical scaleheights that are small compared to the radial extent of the disk \cite{Kenyon:1987kx}. As a result, the majority of the mass (residing near to the disk midplane) is concealed from the parent star. Visual optical depths at the midplane, measured along lines directly from the star, typically exceed $10^{6}$, implying that essentially no stellar photons can penetrate these parts directly. Instead, radiation must be scattered downwards from the surface of the disk \citep{van-Zadelhoff:2003zt}. Scattering is therefore of central importance when discussing the transport of stellar radiation in protoplanetary disks. 

The penetration and absorption of X-ray and FUV radiation determines much of the thermodynamical stratification of the upper disk \citep[e.g.][and references therein]{Aresu:2011tg}. For simplicity we can identify three layers; the uppermost layer being a photodissociation layer illuminated directly by the intense, unattenuated stellar radiation field - this layer is dominated by atoms, since the photodissociation timescale of molecules is very short.  As unscattered photons penetrate into the disk they eventually reach an irradiation surface, defined to be where the optical depth integrated from the star is of order unity, $\tau^*\approx1$ \citep{Calvet:1991pb,Chiang:1997kl,Watanabe:2008lo}.  It is here that, in an average sense, photons undergo their first interactions with the disk.  In the case of interactions with dust grains, this results in attenuation through both absorption and scattering; the latter leading to the emergence of a diffuse component to the radiation field.
Since the radiation field decreases with depth, while the local mass density increases,
we eventually enter an intermediate warm molecular layer, where $\hh$, CO and a host of other
gas-phase molecules may exist in a dynamic equilibrium with the photodissociating FUV field. It is in this warm molecular layer that the photochemical processes are most important. Despite its attenuation, the radiation field is sufficient to maintain grain temperatures above $30$K, thus preventing widespread freeze-out of molecules onto grain surfaces.  At even greater depths we enter the optically thick, dense midplane region containing $>99\%$ of the disk mass.  Here the stellar FUV field is highly attenuated and consists entirely of scattered photons.  With the exception of the innermost few AU, the temperatures may be low enough ($<30$K) to permit the freeze-out of several molecular species \citep{Aikawa:2002hs,Bergin:2007dz}.

The transport of FUV photons is further complicated by the gradual growth and settling of dust grains as the disk evolves towards planetesimal formation \citep{Throop:2001ee,Wilner:2005kn}.   Not only does grain-growth affect the optical properties of individual grains, but the radial and vertical drift of solids relative to gas causes a segregation of the grain-size population \citep{Weidenschilling:1977xw,Dullemond:2004mi}.  The majority of the grain opacity at FUV wavelengths is attributable to relatively small grains with radii $a<1\mu$m, which subsequently reemit absorbed energy at infrared wavelengths, making it possible to estimate their abundance \citep{Dominik:2008hc}.  Observationally, the abundance of small grains (per H nucleus) relative to that found in the ISM, $\epsilon$, typically takes a value much less than one; for example, in the Taurus star-forming region the median value is $\epsilon\sim0.01$ \citep{Furlan:2006fe}.  

Our basic picture of FUV transport is somewhat incomplete once we include $\lya$ photons.  In view of the large cross-section due to resonant scattering by H atoms, $\sigma_{\lya}$, it seems entirely plausible that $\lya$ photons experience their first interactions not with dust grains but with H atoms residing in the uppermost photodissociation layer of the disk (hereafter `H-layer').  The irradiation surface for $\lya$ photons is therefore defined as the location where  $\tau^*=\int \sigma_{\lya} n(H)\approx1$.  The precise location depends quite sensitively on the width of the stellar $\lya$ line profile, although the $\lya$ photons that propagate freely to the disk surface are generally those in the wings of the scattering line profile presented by the disk \citep{Herczeg:2002fu}.   Even hundreds of Doppler widths out in the Lorentz wings of the scattering line profile the cross-section may still be significantly larger than that due to the (possibly settled) dust population.  Figure \ref{fig:voigt} shows the $\lya$ line from TW Hya, plotted next to an estimate of the scattering line profile due to warm H atoms in the disk.   Assuming the photodissociation layer supports H columns of $N^*(H)\ge10^{20}$cm$^{-2}$ (integrated along a line from the star), it seems probable that the $\lya$ field will first be scattered by this atomic H-layer.  The optical depth of dust in this layer will be smaller by a factor of approximately $0.1\epsilon<<1$.  

In stark contrast to $\lya$ transport, FUV-continuum photons will stream unimpeded through the H-layer, eventually striking the disk at the conventional irradiation surface defined by dust.  Here they will begin to scatter, however scattering by typical interstellar dust populations is considered to be somewhat forward-throwing \citep{Gordon:2004lh}.  Multiple scatterings are therefore required before an appreciably diffuse component is generated, and accompanying this scattering will be a degree of pure absorption.  
 

In this paper we explore numerically the basic phenomenology of resonantly scattered $\lya$ in disk models. An emphasis is placed on contrasting the transport of $\lya$ with that of the FUV-continuum field in the vicinity of the $\lya$ line.  We proceed with Section \ref{sec:disks} where we discuss the properties of our disk models.  Radiation transfer of $\lya$ and FUV-continuum photons is dealt with in Section \ref{sec:Radiative-transfer}.  This includes an approximate - but largely self-consistent -computation of the H distribution necessary for $\lya$ transport. Results are presented in Section \ref{sec:Results}.  The paper concludes with a discussion and summary.

\section{Disk models\label{sec:disks}}

\begin{figure}
\begin{centering}
\includegraphics[scale=0.75]{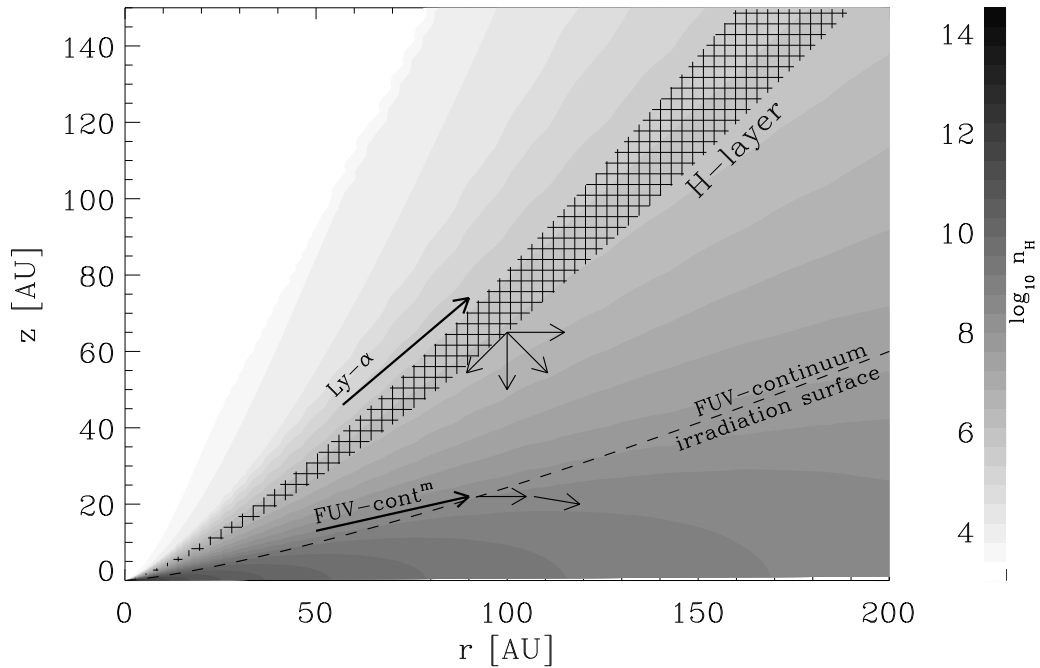}
\caption{The total hydrogenic density distribution, $\nhtot$, in the $\epsilon=0.01$ modified D'Alessio disk (\textit{filled contours}).  A hypothetical atomic hydrogen layer (`H layer') is shown by the hatched region.  Stellar $\lya$ photons strike the upper surface of this layer.  A fraction of this flux is transmitted, emerging isotropically (shown by arrows). In contrast, stellar FUV-continuum photons are shown to strike the irradiation surface determined by dust opacity.  \label{fig:nh_map}}
\end{centering}
\end{figure}

We base our disk structures on a suite of three disk models from \citet{DAlessio:2006rw}.  The  disks are differentiated by the depletion of small grains in their upper layers, $\epsilon=\{0.01,0.1,1.0\}$.  Physically, the parameter $\epsilon$ represents the ratio of dust concentration relative to that found in the interstellar medium.  Although the models include the effects of dust depletion on the hydrostatic disk structure, they lack a separate thermodynamic treatment of the gas, which becomes thermally decoupled at the low densities found in the disk atmosphere.  While the D'Alessio et al. disks suffice for illustrating the essential features of $\lya$ and FUV-continuum transport, a more self-consistent treatment will eventually require the inclusion of detailed coupling between the radiative transfer, thermodynamics  and disk structure.  Several disk models exist that already include the thermal decoupling of gas and dust that generally leads to larger disk scaleheights \citep[e.g.][]{Ercolano:2009oq,Gorti:2009ve,Woitke:2009qq,Owen:2010ly}.  We  assign a nominal FUV luminosity of $0.01L_{\odot}$ to the central star and a kinetic temperature of $T_{g}=1000$K to the gas in the upper layers of the disk.  In view of these modifications we refer to the models as `modified D'Alessio disks'.  A cross-section of the total hydrogen density, $n_H\equiv n(\textrm{H})+n(\hh)$, in one of the disk models is shown in Figure \ref{fig:nh_map}.

The grain-size distribution used in the original D'Alessio calculations follows \citet{Mathis:1977nx}, $dn/da\propto a^{-3.5}$, where $a$ is the grain radius.  In fact there are two populations of grains (`large' and `small') in the D'Alessio models, however the large grains are confined to the disk midplane and play no role in what follows.  For the `small' grain population the grain-size limits are $a_{min}=0.01\mu$m and $a_{max}=0.25\mu$m (double check this).  These limits are consistent with those proposed by \citet{Mathis:1977nx} for interstellar grains.  

\begin{figure}
\begin{centering}
\includegraphics[scale=0.85]{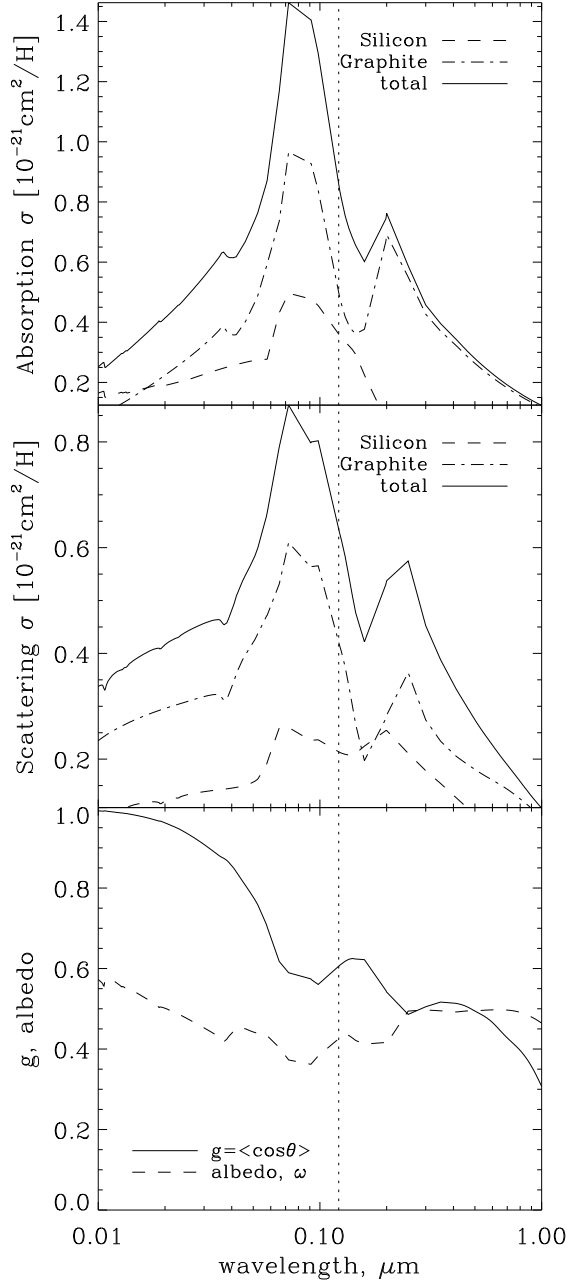}
\caption{Optical properties of the grain ensemble composed of silicon and graphite components (ref to D'Alessio).  \textit{Top} - absorption cross-section.  This paper is primarily concerned with dust opacity at the $\lya$ wavelength ($1215.67\AA$, denoted by vertical dotted line).  \textit{Middle} - scattering cross-section. \textit{Bottom} - single-scattering albedo, $\omega$, and asymmetry parameter, $g$.  At $1215.67$\AA       dust has an asymmetry parameter of $g\sim0.6$ and albedo $\omega\sim0.4$.  \label{fig:dust_properties}}
\par
\end{centering}
\end{figure}

\begin{figure}
\begin{centering}
\includegraphics[scale=0.85]{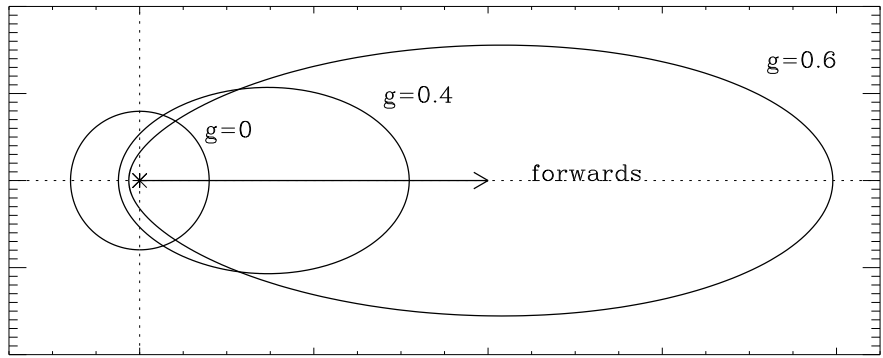}
\caption{The Henyey-Greenstein phase function evaluated for $g=0, 0.4$ and $0.6$ (see Eqn \ref{eq:hg}).  Isotropic scattering has $g=0$ and is a convenient approximation for modeling the resonant scattering of $\lya$ by H atoms.  The dust ensemble used in this paper has an asymmetry parameter of $g\sim0.6$ and is therefore considered to be significantly forward-throwing. \label{fig:hg}}
\par
\end{centering}
\end{figure}

Relevant optical properties of the `small' grain population are shown in Figure \ref{fig:dust_properties}.  In this paper we are primarily concerned with FUV-continuum wavelengths in the vicinity of $1215.67$\AA.  At this wavelength the `small' grain population exhibits an appreciably forward-throwing phase function (asymmetry factor of $g\sim0.6$).  The widely used Henyey-Greenstein phase functions \citep{Henyey:1941bv,Witt:1977gb} evaluated for $g=0,0.4$ and $0.6$ are shown in Figure \ref{fig:hg}.

The computational domain used for the subsequent radiative transfer simulations comprises the modified D'Alessio disks immersed in a background mesh.  The modified D'Alessio disks are spatially 1+1D datasets; the purpose of the background grid of nodes is to transform these models into true 2D distributions.  The resulting discretization consists of a list of nodes, each of which has associated with it a location as well as relevant physical quantities (e.g. densities) which are interpolated from the original D'Alessio models.  The background mesh supports unstructured distributions of nodes in order to accommodate spatially irregular datasets such as the D'Alessio models.  To the extent possible, the initial background grid exhibits grid refinement such that its spatial resolution matches that of the D'Alessio data.

\begin{figure}
\begin{centering}
\includegraphics[scale=0.85]{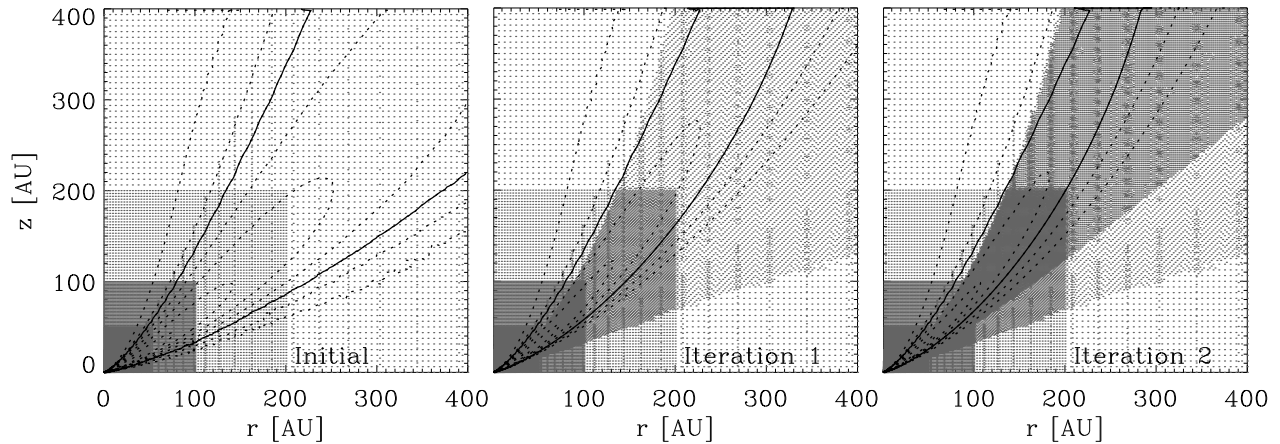}
\caption{Adaptation of the unstructured grid during the iterative H/H$_2$ calculation in the $\epsilon=0.01$ modified D'Alessio disk. \textit{Left} - starting grid comprised of nested regular background arrays and the original D'Alessio 1+1D data.  Total number of nodes is $\sim4\times10^4$.  \textit{Middle} - after one iteration new nodes have been inserted into regions where a change in the H$_2$ molecular fraction has been detected. \textit{Middle} - the node distribution after two iterations.  The region requiring refinement becomes smaller as the H/H$_2$ transition becomes better defined. The H/H$_2$ distribution typically converges to a solution after about 15 iterations, at which point the total number of nodes is $\sim3\times10^5$.  In the inner disk the spatial resolution of the discretization is on the order of \label{fig:grid_adaptive}}
\par
\end{centering}
\end{figure}

\begin{figure}
\begin{centering}
\includegraphics[scale=0.75]{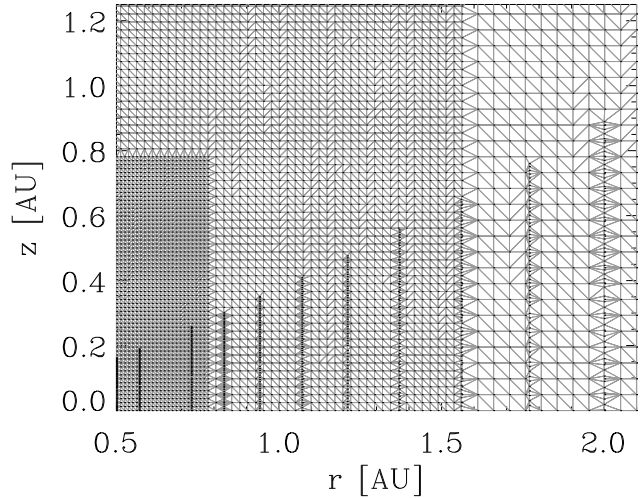}
\caption{Detail of the inner regions ($r\sim1$AU) of the discretized $\epsilon=0.01$ modified D'Alessio disk prior to the H/H$_2$ calculation. Shown are the actual connections between nodes calculated by performing a Delaunay tessellation on the node list.  The resulting triangles are the cells through which photon packets propagate.  As in Figure \ref{fig:grid_adaptive}, the regular array of nodes is the background grid, employing a divide-by-2 scheme to increase the resolution nearer to the star. Note that although these nodes are distributed in a rectangular array, the resulting cells generated by the Delaunay tessellation are triangular.  The original data from the 1+1D D'Alessio disk model are identified as the vertical columns of nodes disrupting the regularity of the 2D background lattice.  In a sense, the Delaunay tessellation heals these discordances, merging disparate sets of nodes. \label{fig:grid_inner}}

\par\end{centering}
\end{figure}

The nodes (constrained to lie in the $r-z$ plane in a cylindrical coordinate system) ultimately form the vertices of triangular cells.  Figures \ref{fig:grid_adaptive} and \ref{fig:grid_inner} give the reader an impression of the spatial distribution of nodes and how they result in a tessellation of triangular cells.  The triangular cells are a result of the connectivity obtained by performing a Delaunay tessellation on the node list \footnote{See, for example, http://www.qhull.org/}.  In 2D the triangle is the simplex: the simplest shape that can tessellate the volume \citep{Okabe:1992dz}.  In 3D the simplex is the tetrahedron.  There are many alternative ways to connect the nodes that give rise to a sensible tessellation of triangular cells, however the Delaunay tessellation has unique properties that arguably yield the optimal connectivity.  First, the connections link those nodes that compete for space.  This is most easily understood by considering the corresponding Voronoi diagram (the dual of the Delaunay tessellation).  Second, the resulting triangles maximize the minimum internal angle, thus reducing the frequency of skinny triangles.  While the meshes in Figs. \ref{fig:grid_adaptive} and \ref{fig:grid_inner} ostensibly appear regular, they are treated as fundamentally unstructured.  Had the background mesh been cast randomly (e.g. according to a Poisson point process) it would be possible to shed all traces of the underlying coordinate system in the spatial discretization \citep{Ritzerveld:2006fu}. For these reasons the Delaunay tessellation is becoming an attractive choice for generating high-quality, unstructured discretizations of spatial objects.  

\section{Radiative transfer\label{sec:Radiative-transfer}}

Very few radiative transfer problems of astrophysical interest can be solved analytically without resorting to significant simplifications \citep{Chandrasekhar:1960kl}.   The transport of radiation in protoplanetary disks is no exception. In view of the geometrical complexity and assortment of opacity sources, we opt to solve the transport problem using the highly versatile Monte Carlo simulation method.  Numerical details of our Monte Carlo radiative transfer code are described in the Appendices.  In this section we restrict the discussion to aspects which are relevant to the physical context.

\subsection{Transport of FUV-continuum}

The transport of FUV-continuum photons is assumed to be governed solely by interactions with dust grains.  At extreme levels of dust settling ($\epsilon<0.001$) it is possible that Rayleigh scattering by $\hh$ molecules becomes relatively important. Anisotropic scattering by grains is treated using the one-parameter Henyey-Greenstein phase function, the angular scattering probability distribution of which is given by,

\begin{equation}
p(\theta)=\frac{1}{4\pi}\frac{1-g^2}{[1+g^2-2g\cos\theta]^{3/2}}.\label{eq:hg}
\end{equation}
where $g=[-1,1]$ is the anisotropy parameter.  Isotropic scattering corresponds to $g=0$, forwards scattering to $g=1$, and backwards scattering to $g=-1$ (Figure \ref{fig:hg}).   The single-scattering albedo $\omega\sim0.4$ indicates that these grains have a slight preference towards absorption rather than scattering.  From $\omega$ and $g$ we can conclude that multiple scatterings will be required in order to generate a diffuse field that can penetrate the disk vertically.  Accompanying these scatterings will be significant absorption.  

\subsection{Transport of $\lya$.  The $\textrm{{H}/}\textrm{\textrm{H}}_{2}$ distribution\label{sec:Disk-H}}

The transport of $\lya$ is treated like FUV-continuum but with the addition of resonant scattering.  Accordingly,  it is necessary to first determine the distribution of atomic hydrogen.  The formation of $\hh$ in dense, dusty environments usually occurs on the surfaces of dust grains \citep{Savage:1977ij,Spitzer:1978fv,Cuppen:2005bs}.  The destruction of $\hh$ occurs with $\sim10$\% probability following an absorption of a FUV photon into the Lyman-Werner band system \citep[$912-1100\textrm{\AA}$,][]{Hollenbach:1999qf}. Since the photodissociation of $\hh$ occurs through line absorptions, these transitions become optically thick over very modest columns, $N(\hh)\sim 10^{14}$cm$^{-2}$. Beyond this point we see a run-away formation of `self-shielded' $\hh$. While any photo-destruction dominated process that competes with formation can in principle result in self-shielding \citep{Bethell:2009uq}, the concentration of $\hh$ opacity in discrete lines makes it readily self-shielding. Balancing formation and destruction rates gives the steady-state molecular fraction \citep{Spaans:1997kx},
\begin{equation}
\fhh\equiv\frac{\nhh}{\nhtot}=\frac{\nhtot\rh\epsilon}{\zeta+2\nhtot\rh\epsilon},\label{eq:H2}\end{equation}
where $\rh$ is the high temperature $\hh$ formation rate coefficient \citep{Cazaux:2004fu,Cazaux:2010dk}, and $\zeta$ is the $\hh$ photodissociation rate $\zeta=\zeta_{0}F_{s}[\colhh]$. For the self-shielding function, $F_{s}[\colhh]$, we adopt the closed-form expression from \citet{Draine:1996oz}, although comparison simulations were also performed using the tabulated results of \citet{Lee:1996tw}. These approximations were shown to result in H-layers with comparable geometries and column densities.  In the context of our Monte Carlo simulation we evaluate the self-shielding function along the length of each photon packet trajectory \cite[e.g.][]{Spaans:1997kl}.   The inclusion of a full treatment of $\hh$ photodissociation is currently beyond the capabilities of most multidimensional photochemical codes \citep{Rollig:2007il}. 

\subsubsection{Iterative $\textrm{{H}/}\textrm{\textrm{H}}_{2}$ calculation}

The calculation of the H/$\hh$ distribution is an iterative one, since to compute $\nhh$ at a point requires us to know the degree of self-shielding due to $\nhh$ everywhere else. An initial distribution of H and H$_2$ is required to start the iterative calculation.  From this our $\hh$ will `grow' as self-shielding is established over the course of repeated iterations.  Ideally, this initial condition should be constructed to underestimate the $\hh$ density: if $\hh$ is overestimated then regions that are artificially self-shielded will be slowly photodissociated away over the course of the iterative process.  In such over-shielded cases, a layer of excess H$_2$ with a thickness corresponding approximately to the self-shielding column ($N(\hh)\sim10^{14}$cm$^{-2}$) will be photodissociated in each iteration.  If the excess $\hh$ is distributed over a large volume, the convergence towards the true H$_2$ distribution will be very slow.  Since the introduction of self-shielding always \textit{increases} the H$_2$ density, we can set a strict lower limit by computing $\nhh$ using Eqn \ref{eq:H2} ignoring  all forms of shielding (this includes dust attenuation, although we retain the geometrical inverse-square dilution of radiation).  The resulting $\nh$ is shown in the left-hand panel in Figure \ref{fig:grid_adaptive}.  Even without self-sheilding, atomic hydrogen is limited to the upper disk layers due solely to the increase in density that promotes H$_2$ formation deep in the disk.  

The iterative process commences with a full radiative transfer simulation at a single representative Lyman-Werner band wavelength (centered at $\lambda=1000\textrm{\AA}$).   The accumulation of H$_2$ column density along photon packet trajectories is recorded, and information regarding the self-shielded radiation field is deposited in each cell traversed by the photon packet. Dust scattering and absorption are both included. However, since $\hh$ shielding proceeds so vigorously, dust shielding plays a relatively minor role in determining the final H/$\hh$ distribution.  This is especially true in dust-settled scenarios, although a rigorous consideration must establish the criteria that separates purely $\hh$ self-shielding situations from those in which $\hh$ self-shielding is initiated by dust attenuation \citep{Wolfire:1995vn}.  

Once approximately $10^6$ photon packets have been run, the resulting self-shielded radiation field is used to compute a new $\nhh$ using Eqn \ref{eq:H2}.  Since the original grid is generally too coarse to properly resolve the emerging H/H$_2$ transition we refine the grid before proceeding with a new iteration.  If the $\hh$ distribution is described on grid that is too coarse, individual cells may self-shield, causing a local overestimation of $n(\hh)$ that may subsequently propagate through the grid.  The refinement criterion is based upon the fractional change in $\nhh$  between iterations.  This ensures that the spatial resolution of the grid is increased in regions which are most sensitive to changes in $\nhh$.    Regions that are either entirely unshielded or totally self-shielded ($\nhh\rightarrow0.5\nhtot$) will not receive grid refinement.  The refinement of a cell is achieved  by placing a single node in the geometric center of the cell.  Once new nodes have been added a Delaunay tessellation is performed on the augmented node list, resulting a new set of cells.  Physical quantities are interpolated onto this new grid and the next iteration cycle is begun.

The criteria for a converged solution must not only require a solution that does not change significantly with further iterations, but also that every cell in the H/$\hh$ transition must not contain a sufficiently large $\colhh$ to self-shield all by itself.  For a cell to be considered part of the H/$\hh$ transition it must exhibit a $\nhh$ that is greater than its unshielded value and less than its maximal possible value of $0.5\nhtot$.

The nodes and $\nh$ distributions after the first and second iterations in the $\epsilon=0.01$ model are shown in Figure \ref{fig:grid_adaptive}.  It is worth noting that the uppermost layers are entirely photodissociated so that $\nh\approx\nhtot$ throughout the iterative procedure.  It is the \textit{lower} boundary of the H layer that is affected by self-shielding, and this can be seen to rise in height as the bulk of the disk gradually becomes molecular.  For the modified D'Alessio disks we typically undertake 15-20 iterations.  Convergence is assumed when $\nhh$ changes by less than 5\% everywhere in the domain.  It is important to realize that Monte Carlo noise will generate variance between iterations, even after many iterations.  If the Monte Carlo noise level is known then Eqn \ref{eq:H2} can be used to estimate its effect on $\nhh$.  

To summarize, the basic steps of the iterative H$_2$ calculation are as follows;
\begin{enumerate}
\item Compute an initial distribution (lower limit) of H$_2$ due to dissociation by the unattenuated stellar field (i.e. ignoring $\hh$ self-shielding and dust attenuation).
\item Propagate photon packets using Monte Carlo methods, following $\hh$ self-shielding and dust-absorption along trajectories.
\item Compute new H$_2$ distribution due to this photodissociation field.
\item Add nodes to regions where the H/H$_2$ transition is detected.   Interpolate physical properties onto the new mesh.
\item Repeat from step 2 until solution has converged.
\end{enumerate}

\subsubsection{$\lya$ transport}

\begin{figure}
\begin{centering}
\includegraphics[scale=0.75]{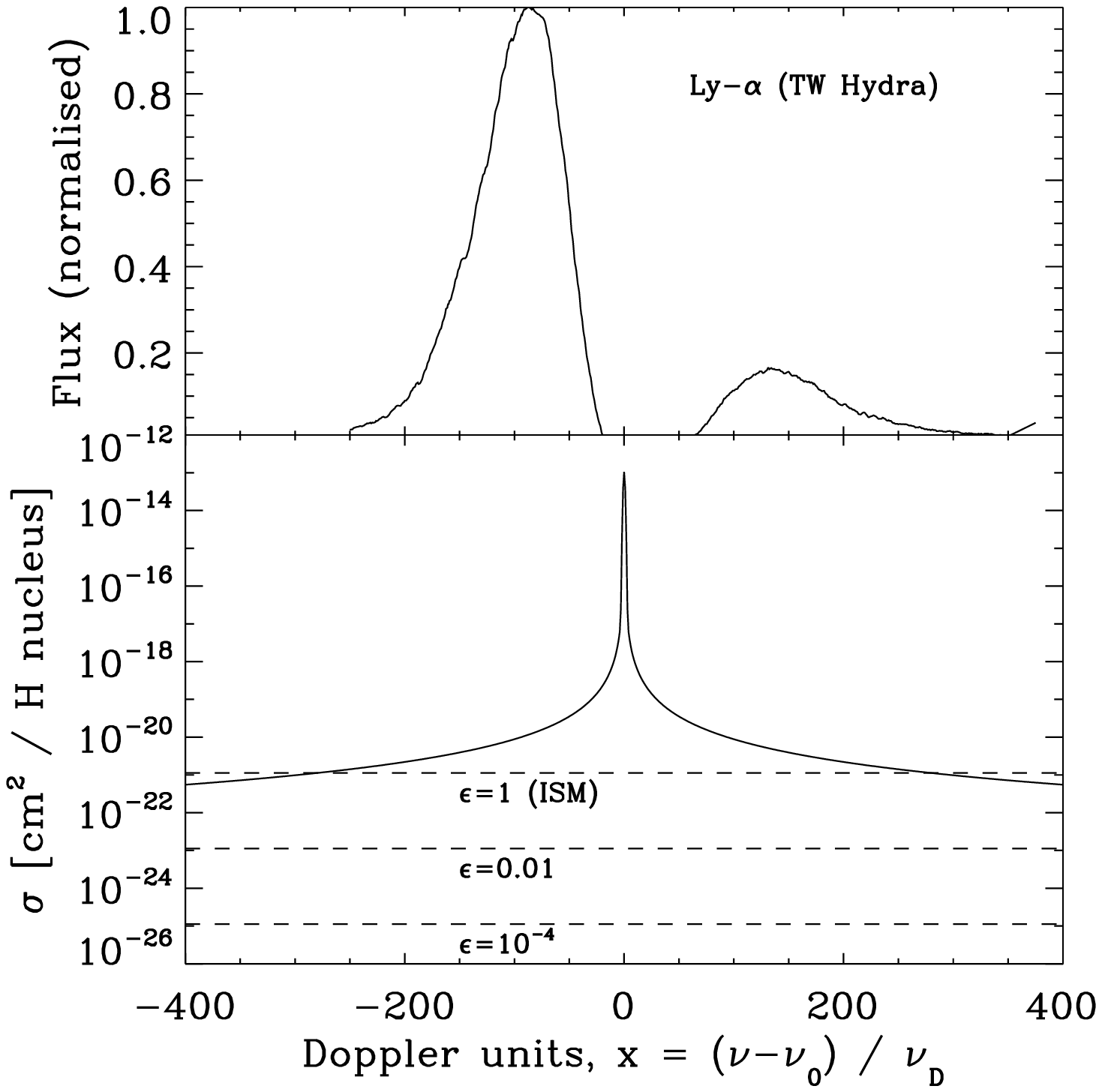}
\caption{\textit{Top} - $\lya$ spectrum of TW Hya expressed in Doppler units,
$x$, relative to line center \citep{Herczeg:2002fu}.  \textit{Bottom} - The resonant scattering Voigt profile
(\textit{solid line}) presented by a gas of H atoms with kinetic temperature $T=1000$K.  The Lorentzian wings dominate the profile at displacements of $|x| \ge 3$.  Dust opacity (\textit{dashed line}) is essentially constant over this small wavelength range, scaling in proportion to the dust abundance $\epsilon.$ \label{fig:voigt}}
\end{centering}
\end{figure}

Once the H-layer has been calculated it is possible to propagate $\lya$ photons.  The resonant scattering process is simplified somewhat by associating with it a Voigt line profile, $\phi(x)$, and an isotropic phase function. \citet{Ahn:2001ys} and \citet{Laursen:2007ly} discuss the treatment of resonant $n=1\rightarrow 2$ transitions in more detail, although for our purposes the simplified treatment is sufficient \footnote{A more complete microphysical treatment is required if polarization is to included.}. As discussed previously, most $\lya$ photons from the TW Hya spectrum scatter in the Lorentzian wings of the Voigt profile (see Figure \ref{fig:voigt}).  Unlike the Doppler core, which is relatively temperature sensitive, the wings of the scattering profile do not change greatly with the temperature of the gas.  This somewhat mitigates the ad hoc temperature imposed upon the gas in our modified D'Alessio disks.
 
 The convolution of the Gaussian (thermal broadening) and Lorentz (natural) line profiles gives rise to the Voigt profile,

\begin{equation}
\sigma(a,x)=f_{12}\frac{e^2\sqrt{\pi}}{m_ec\Delta\nu_D}H(a,x),
\end{equation}
where $f_{12}$ is the oscillator strength for the the $1\rightarrow2$ transition, $e$ is the electron charge, $c$ the speed of light, and $m_e$ the electron mass. $\Delta\nu_D\equiv \nu_0 v_{th}/c$ is the thermal Doppler width, where $v_{th}$ is the mean thermal velocity of the scatterers and $\nu_0$ is the frequency at line center. $H(a,x)$ is the Voigt profile defined by the integral,

\begin{equation}
H(a,x)=\frac{a}{\pi}\int_{-\infty}^{+\infty}\frac{\exp{-y^2}}{(x+y)^2+a^2}dy.
\end{equation}


While the scattering is coherent in the frame of the scattering atom, an overall frequency change occurs due to the difference between the incoming and outgoing Doppler shifts required to change between atom and disk frames.  This is known as partial frequency redistribution \citep{Hummer:1962qo}. The fractional change in photon frequency is of order the thermal Doppler width $\Delta\nu_D\equiv \nu_0 v_{th}/c$ where $v_{th}$ is the mean thermal velocity of the scatterers and $\nu_0$ is the frequency at line center.  Partial redistribution of the scattered $\lya$ photon is included in our code using microscopic Monte Carlo techniques based on \citet{Avery:1968fy} and \citet{Zheng:2002mb}.  However, since the intrinsic stellar $\lya$ line shape is already very broad ($\sim300$ Doppler widths, see Fig. \ref{fig:voigt}), and the H-layer not extremely optically thick, little appreciable frequency redistribution of the $\lya$ line is expected. As such one can largely ignore frequency redistribution in our models, and instead treat the resonant scattering as elastic.  This simplification greatly expedites the calculation.  We ignore systematic Doppler shifts due to the azimuthal velocity component of the disk \citep[$\sim30$kms$^{-1}$ at 1AU,][]{Gayley:2001dq} since for much of the disk this induces Doppler shifts that are modest compared to the stellar $\lya$ line width ($>300$kms$^{-1}$).


\section{Results\label{sec:Results}}

The vertical distribution of hydrogen in the three disk models is shown in Figure
\ref{fig:columns}. The molecular hydrogen column density $\colhh$ shows the characteristic
S-shape associated with the onset of self-shielding as we descend
into the disk. The vertical column density of atomic hydrogen clearly varies with the degree of dust depletion,
$\colh\sim10^{19}-10^{21}\textrm{cm}^{-2}$ - this is the column thickness of
the H-layer that scatters $\lya$ photons. Thicker H-layers are associated with dust depleted models, due to the proportional reduction in the $H_2$ formation rate.  The precipitous drop in $\nh$ seen in Figure \ref{fig:columns} is indicative of the sudden onset of $\hh$ self-shielding.  The \textit{vertical} optical depth (for resonant scattering) of the H-layer depends upon the
wavelength displacement from line center; for the $\lya$ spectrum shown in Figure \ref{fig:voigt} the photon-averaged optical depths are $\tau_{\lya}^{z}\sim0.2,1,6$ for $\epsilon=1,0.1,0.01$, respectively.
Although these vertical optical depths are not large, when viewed from the star the \textit{slant} optical depth this H-layer is typically $\tau_{\lya}^{*}\sim20\tau_{\lya}^{z}\gg1$ and thus \textit{the H-layer will intercept essentially every stellar $\lya$ photon incident upon it}.  We are mostly interested in following the portion of $\lya$ emerging from the lower surface of the H-layer. The remainder is scattered into space from the upper surface.  It is worth noting
that the H-layer is not so optically thick to these line-wing photons that they become line-trapped and destroyed by dust absorption \citep{Neufeld:1990os}.  While these results suggest that resonant scattering of $\lya$ is indeed important, the H-layers computed lack the self-consistency required to render them entirely realistic.  

\begin{figure}

\centering{}
\includegraphics[width=0.8\textwidth]{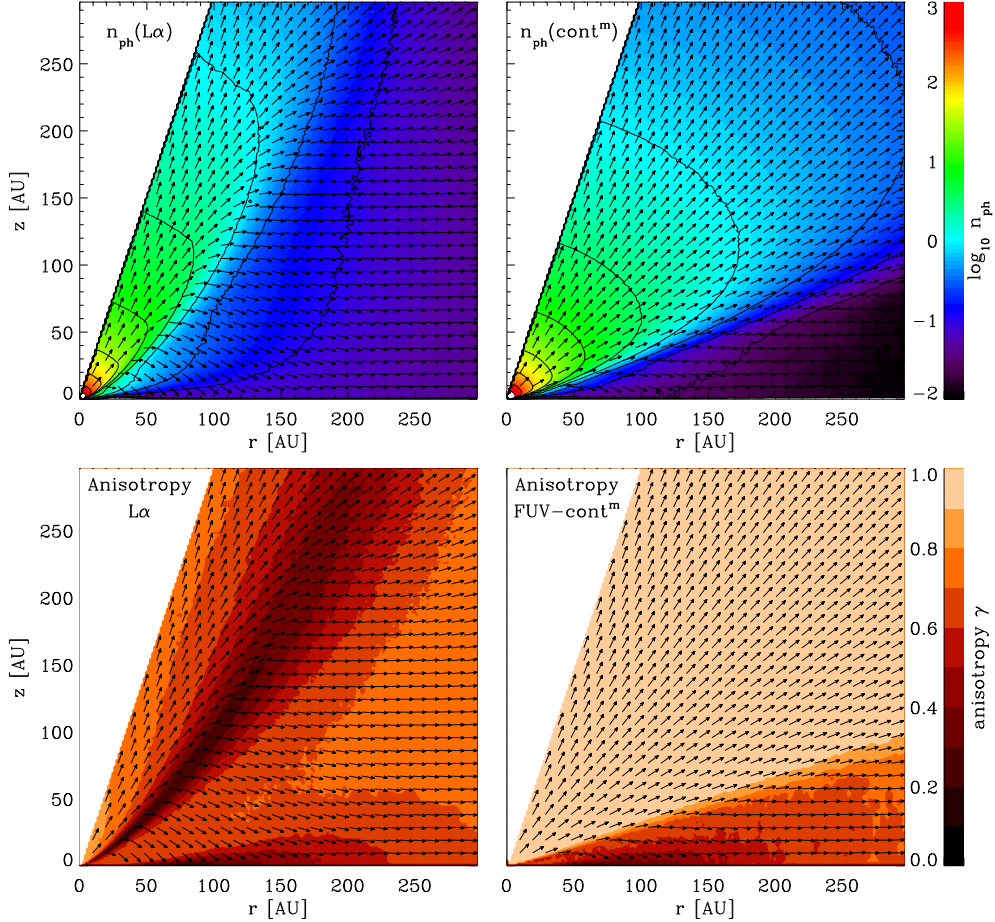}
\caption{Maps ($r-z$) of the photon density and anisotropy of the $\lya$
(\textit{left-hand panels}) and FUV-continuum (\textit{right-hand
panels}) radiation fields in the $\epsilon=0.01$ modified D'Alessio
disk model. The star is located at the origin. \textit{Top} - photon density
(units $\textrm{cm}^{-3}$). Black arrows indicate the direction of
the net flux. \textit{Bottom} - the anisotropy of the radiation field,
$\gamma\equiv\mid\int I({\bf k}){\bf k}d\Omega\mid/4\pi J$, where
$\mathbf{k}$ is the unit direction vector, $I({\bf k})$ is the intensity, and $J$ is the mean intensity. The limit $\gamma=1$
implies unidirectionality and $\gamma=0$ implies isotropy. The flared
feature of low anisotropy seen in the $\lya$ panel (\textit{lower
left}) is attributable to (and coincident with) the resonantly scattering H-layer. Note also how the net $\lya$ flux
emerging from the base of the H-layer does so perpendicularly. \label{fig:maps}}
 
\end{figure}

Figure \ref{fig:maps} shows spatial maps providing a visual comparison of the $\lya$ and FUV-continuum
photon densities in the $\epsilon=0.01$ modified D'Alessio disk. In addition to the photon density $n_{ph}$
we also show flux arrows (direction only), and the anisotropy of the field $\gamma$ (the ratio of net flux to photon density).  The highly flared H-layer and its isotropizing effect on the $\lya$ field is clearly seen in the left-hand anisotropy panel.  The net flux of transmitted $\lya$ photons emerges from the H-layer perpendicularly, providing a more direct illumination of the disk below.  

\begin{figure}
\begin{centering}
\includegraphics[width=0.65\textwidth]{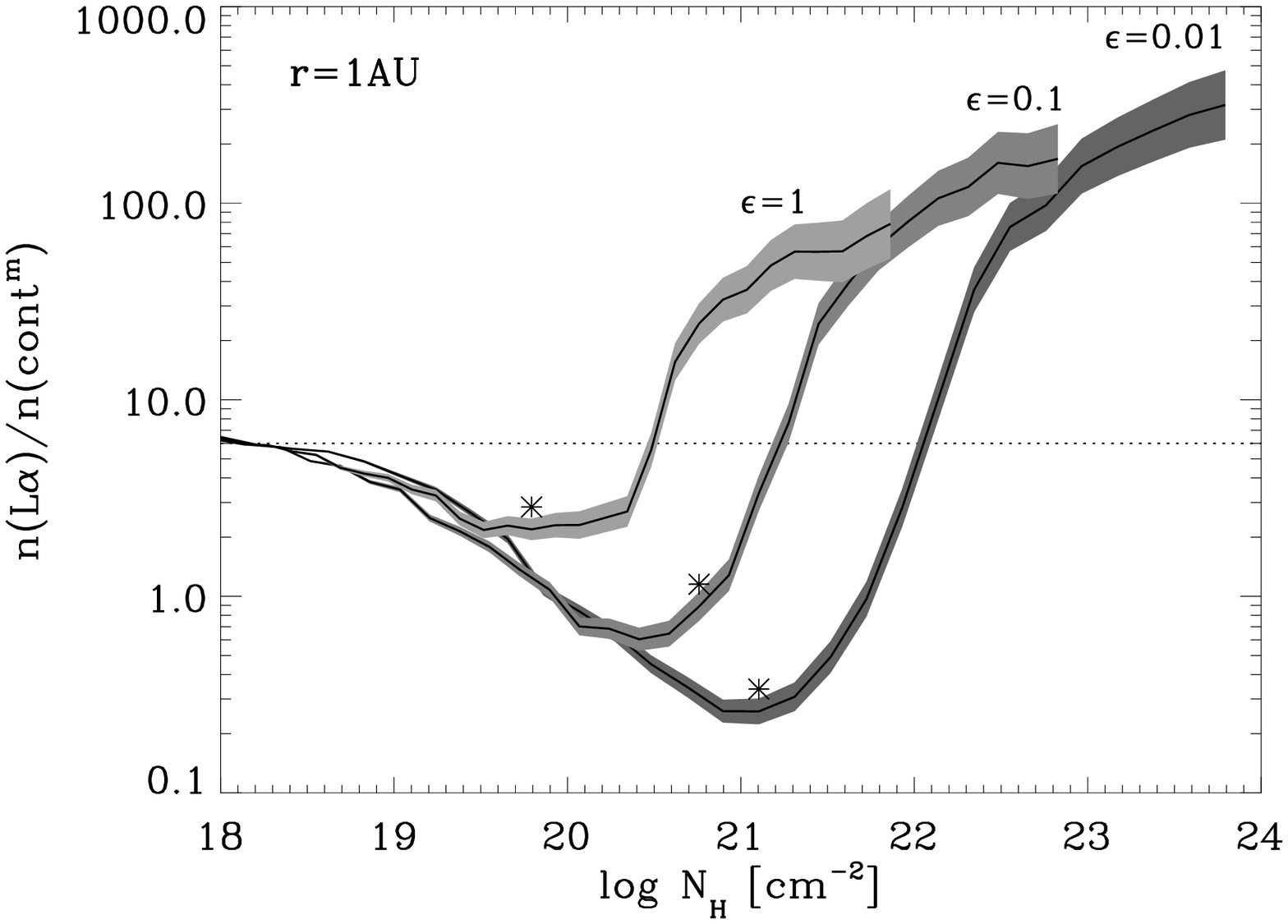}
\par\end{centering}

\begin{centering}
\includegraphics[width=0.65\textwidth]{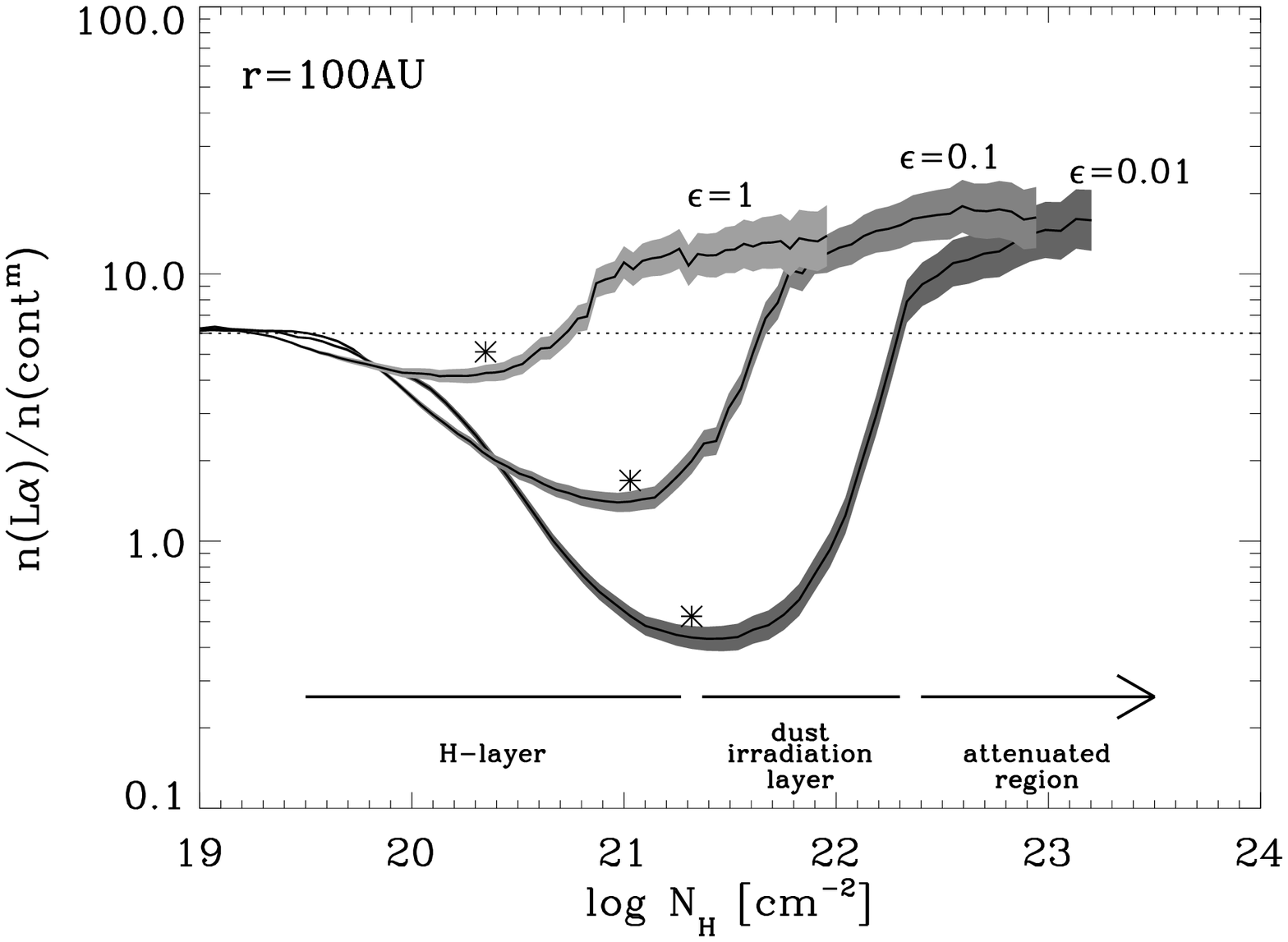}
\par\end{centering}

\caption{The ratio of $\lya$ and FUV-continuum photon densities as a function of vertical column density in the modified D'Alessio disks. \textit{Top} - at $r=1$AU the ratio departs from the intrinsic stellar ratio ($\sim$6) as we descend into the disk (increasing $\hcol$).  The asterisks denote the lower surface of the H-layer (c.f. Fig. \ref{fig:columns}). \textit{Bottom} - the same at $r=100$AU. The shaded areas provide an estimate of the Monte Carlo noise. The curves are terminated at the depth where the noise begins to degrade the data significantly. Interpretative text has been added for the $r=100$AU, $\epsilon=0.01$ case. \label{fig:ratios}}

\end{figure}

Vertical profiles of the $\lya/$FUV-continuum photon density ratio are shown in Figure \ref{fig:ratios} for $\epsilon=0.01,0.1,$ and $1$, at radii $r=1$ and $100$AU. The same qualitative behavior of the ratio is seen at all radii, although the quantitative effect is greatest in the inner disk. It is clear from Figure \ref{fig:ratios} that the H-layer is FUV-continuum dominated, whereas the molecular disk (and therefore the vast majority of disk mass) is $\lya$-dominated.  Ultimately, the enhancement of the $\lya$ photon density (above the intrinsic stellar value) may exceed an order of magnitude.  The ratios eventually asymptote to a constant value deep into the disk where both fields are approaching the diffusive limit and therefore behaving similarly.

\section{Discussion\label{sec:Discussion}}

Regardless of the value of $\epsilon$, essentially every stellar $\lya$ photon is intercepted by the highly flared H-layer.
Upon the first scattering the $\lya$ field is completely isotropized and a fraction $(<50\%)$is transmitted through to the base H-layer. This is clearly seen in Fig. \ref{fig:maps} (top left and bottom left panels).  Viewed from within the molecular region the $\lya$ would seem to
originate in a diffuse blanket overlying the disk, analogous to the diffuse
transport of sunlight on an overcast day.  This is similar to the penetrative advantage that interstellar photons (bathing the disk isotropically) would have \citep[e.g.][]{Willacy:2000zr}, however in our models the stellar $\lya$ field is typically orders of magnitude more intense.  Once in the realm of the molecular disk, $\lya$ transport will be controlled primarily by dust.  

FUV-continuum photons pass unimpeded through the H-layer until they reach the dust
irradiation surface. Until this point the field only experiences geometric inverse-square law diminution - in
contrast to the $\lya$, which has been additionally processed by the H-layer. In this regime the $\lya$/FUV-continuum photon density
ratio favors the unimpeded FUV-continuum photons. The FUV-continuum eventually impinges upon the dust irradiation
surface, doing so at very small angles ($\leq0.05$ degrees).  The relative inefficacy of scattering by dust grains is exacerbated by accompanying absorption. It is straightforward to show with 1D planar radiative transfer models that the energy density
transmitted downwards from this dust irradiation surface is typically of order $\leq1\%$ of
that incident upon it \citep{Chandrasekhar:1960kl}. Consequently, the $\lya$/FUV-continuum photon
density ratio undergoes a dramatic reversal in the vicinity of the dust irradiation surface, resulting in a $\lya$-dominated
field.  This resurgence appears to more than compensate for the reduction of $\lya$ density due to the H-layer.  In the asymptotic limit (large optical depth) of the analytic planar model both fields are attenuated
according to $n_{ph}\propto\exp\left[-k\tau_{z}\right],$where $\tau_{z}$
is the vertical optical depth due to dust, and $k$ is a decay constant
that depends only on the scattering properties $(\omega,g)$ of
dust \citep{Flannery:1980xr}.  As observed, the $\lya$/FUV-continuum ratio approaches a constant value deep inside the disk (the `attenuated region' in Figure \ref{fig:ratios}).  At this point both $\lya$ and FUV-continuum fields are largely isotropized and their transport controlled by dust alone.  In effect, the contrasting early life-histories are imprinted into boundary conditions that determine the proportionality coefficients in the asymptotic regime.  A simplified diagrammatical representation of the transport routes is shown in Figure \ref{fig:nh_map}. 

As mentioned previously, the D'Alessio disks to not include separate thermodynamic treatments for gas and dust.  In effect they assume that the two components are thermally coupled.  Models show that this coupling breaks down at low densities, causing the gas temperature to rise above that of dust \citep{Glassgold:2004ec}.  In turn this increases the local scale height of the gas while decreasing its density.    Although not explored in this paper, the thermal decoupling will most likely increase the extent of atomic H in the upper layers of the disk.  This is due to two factors; first, the formation rate of H$_2$ is density dependent and an increase in the disk scaleheight will be accompanied by a decrease in volumetric density.  Second,  H$_2$ formation on grains becomes increasingly inefficient at $T\ge 500K$ due to the short time that H atoms spend on grain surfaces before thermally escaping \citep{Cazaux:2004fu}.


The settling of dust affects the $\lya$
and FUV-continuum fields in two distinct ways. First, the removal
of dust decreases the efficacy of the $\hh$ formation process, resulting in a thicker H-layer (Eqn.
\ref{eq:H2}, Fig. \ref{fig:columns}). This ensures that the
H-layer processes every stellar $\lya$ incident upon it.  Second, the irradiation
surface for FUV-continuum photons recedes into the disk as $\epsilon$ decreases.  The disk becomes flatter and intercepts a
smaller fraction of stellar FUV-continuum photons.  A self-consistent disk model that couples disk structure with our new radiative transfer results is required in order to quantify these effects more precisely.

We conclude with a brief summary of the main results:
\begin{enumerate}
\item The vertical profile of the $\lya$/FUV-continuum photon density ratio is strongly stratified, and may vary by
orders of magnitude relative to the intrinsic stellar ratio. The stratification is strongest in the inner disk.
\item The radiation field in the vicinity of the disk midplane - where most of the disk mass resides - is relatively enhanced in $\lya$.  This enhancement is in addition to that already present in the stellar spectrum.
\item If previous radiative transfer calculations most closely resemble our FUV-continuum results, then the omission of resonantly scattered $\lya$ implies that the mass-averaged intradisk FUV field has been systematically underestimated.  This directly affects estimates of photoelectric heating rates, thermal balance, and the resulting hydrostatic structure.
\item The settling of dust ($\epsilon<1$) lowers the dust irradiation surface and thickens the resonant scattering
H-layer by inhibiting $\hh$ formation.  Both effects contribute to a more stratified  $\lya$/FUV-continuum ratio, ultimately resulting in a greater $\lya$ enhancement in the molecular disk.
\item The presence of an optically thick, high-albedo H-layer implies that the direct observation of $\lya$ scattered by the disk ($\geq50\%$ of incident photons) may be possible in sufficiently close, low extinction systems (e.g. TW Hya). These observations may place constraints on the shape and thickness of the H-layer.
\item A $\lya$-dominated radiation field drives differential photo-chemical processes in both the gas and grain surfaces \citep{Bergin:2003la,Fogel:2011ys}. Numerous pairs of closely-related molecules exhibit a similar differential response to the spectral form of the UV field \citep{vDishoeck:2006kc}. In view of the strongly stratified $\lya$/FUV-continuum ratio one might also expect the photochemical environment to be similarly stratified.
\end{enumerate}


\acknowledgements{}

We thank P. D'Alessio and N. Calvet for providing the disk models.  We also thank G. Herczeg for providing the $\lya$ spectrum for TW Hya. TB expresses gratitude to members of the IDL-pvwave mailing list who provided useful suggestions regarding the computational management of unstructured grids. The authors acknowledge support from NASA grant NHN08 AH23G.

\appendix

\section{Monte Carlo radiative transfer\label{sec:MC}}

Markov Chain Monte Carlo simulation is an intuitive and versatile technique for numerically solving high-dimensional, geometrically complex transport problems.  Although the method permits many degrees of abstraction, the most intuitive implementation simply involves following the stochastic life-histories of many discrete particles (in our case, photons or photon packets) as they travel through a medium, experiencing interactions that are modeled probabilistically \citep{Witt:1977gb,Audic:1993lh,Code:1995fu,Bjorkman:2001pi,Gordon:2001ff,Whitney:2003mi}.  If we consider the basic problem of photons being absorbed and scattered then it is primarily the inclusion of scattering that complicates the calculation, and thus motivates the use of the Monte Carlo method.  Additionally, the Monte Carlo method places essentially no constraints on the geometry involved, provided a meaningful spatial discretization can be obtained.  

Our Monte Carlo simulation method is conceptually similar to \citet{Witt:1996cr}.  The reader is referred to this paper for a description of the fundamental steps, which we briefly enumerate here.   

\begin{enumerate}
\item The $i^\textrm{th}$ stellar photon packet (from a total of $N_{ph}$) starts at $r=z=0$ with a luminosity $W_i=1/N_{ph}$ and direction $\cos\alpha=i/N_{ph}$, where $\alpha$ is its elevation angle above the midplane.  This is recast into a cartesian direction vector $\mathbf{k}=(k_x, k_y, k_z)$ where $k_x=\cos\alpha$, $k_y=0$, and $k_z=\sin\alpha$. The star is an isotropic light source with total luminosity of $L_*=\sum_{i}W_i=1$.
\item Randomly sampling Beer's Law generates the optical depth to the next  scattering event, $\tau_{scat}=-\ln(1-p)$, where $p$ is a uniformly distributed random number in the range $[0,1]$.  
\item The photon trajectory is propagated through the medium until it has traversed an optical distance $\tau_{scat}=\int \kappa \textrm{d}s$, where the opacity $\kappa=\sigma_{dust}^{scat} \nhtot$ for FUV-continuum photons and $\kappa=\sigma_{dust}^{scat} \nhtot + \sigma_{\lya} n(H)$ for $\lya$ photons.  The photon is transported from cell-face to cell-face (see Appendix \ref{sec:intercept}), each time traversing some distance $\Delta s$ and optical distance $\kappa \Delta s$, until the scattering location is overstepped, at which point the photon is regressed back to the exact scattering location.  The optical properties within a cell are considered to be homogeneous.  Although the photon is traveling in 3D space, at each face-intercept  its coordinates and direction vector are rotated back into the $r-z$ plane.  Relevant information (e.g. packet luminosity $W_i$) is deposited as the photon passes through each cell.  There is a continual attrition of photons due to absorption, causing a decrease in the weight-factor $W_i\rightarrow \exp[(\omega-1)\omega^{-1}\tau_{scat}] W_i$, where $\omega$ is the albedo.
\item The scattered photon direction is found by performing these steps; transforming into a frame aligned with the photon direction vector, finding the scattering angle by randomly sampling the scattering phase function, and finally transforming back into the coordinate frame of the disk.  This coordinate transformation is an application of Euler angles.  In the case of $\lya$, the type of scattering must first be resolved.  If a uniformly distributed random variate $p\in[0,1]$ satisfies $p<\sigma_{\lya} \nh/(\sigma_{dust}^{scat} \nhtot + \sigma_{\lya} \nh)$, then we execute resonant scattering, otherwise the scattering is by dust. 
\item Repeat from step 2 until photon leaves domain.

\end{enumerate}

The Henyey-Greenstein phase function used for dust scattering has a number of convenient features that make it attractive for use in Monte Carlo simulations.  Not only can its probability distribution (Eqn \ref{eq:hg}) be integrated over angle to produce a closed-form cumulative probability distribution, but the result can be algebraically inverted, permitting the expression of the scattering angle, $\mu \equiv \cos \theta$, in terms of a uniformly distributed random number, $p\in[0,1]$,

\begin{equation}
\mu=\frac{1}{2g}\left[1+g^2-\left(\frac{1-g^2}{1+g(2p-1)}\right)^2 \right].\label{eq:hginv}
\end{equation}

We approximate $\lya$ scattering as isotropic.  This simplifies matters considerably, since the new scattered direction $\mathbf{k}$ is simply the result of randomly sampling directions over the unit sphere, thus avoiding the computationally expensive Euler angle calculation.

We restrict our computational domain to $z\ge0$ i.e. the volume above the disk midplane.  This is equivalent to asserting a mirror plane at $z=0$; if any photons reach the disk midplane they are reflected $k_z\rightarrow-k_z$.    The solution is now smooth and symmetric about the disk midplane, properly accounting for photons that pass vertically from one side of the disk to the other.   When projected onto the $r-z$ plane, straight lines typically appear as hyperbolae. Photons may escape by reaching the outer and upper edges of the domain.  In fact, the Delaunay tessellation produces a convex domain perimeter, which in general is not restricted to a rectangular form.\footnote{The outer perimeter  - or `convex hull' - is the outline that would be formed by a taught string enclosing the entire set of nodes.}.  







\section{Intersection between photon path and cell walls\label{sec:intercept}}

Although the unstructured grid is specified in 2D,  we still need to envision the cells as 3D entities in order to use them in the Monte Carlo radiative transfer simulation.  This is because a photon travels in 3D (it is not constrained to a single $r-z$ plane).  By employing an axisymmetric cylindrical framework we are asserting invariance upon rotation about the $z$ axis.  Once rotated, the triangular simplex is extruded into a torus with a triangular cross-section.   Photon trajectories will be propagated through these volumes, requiring geometric tools that allow us to keep track of the intercepts between the photon path and the surfaces of  these contiguous volumes.  

\begin{figure}
\begin{centering}
\includegraphics[width=0.65\textwidth]{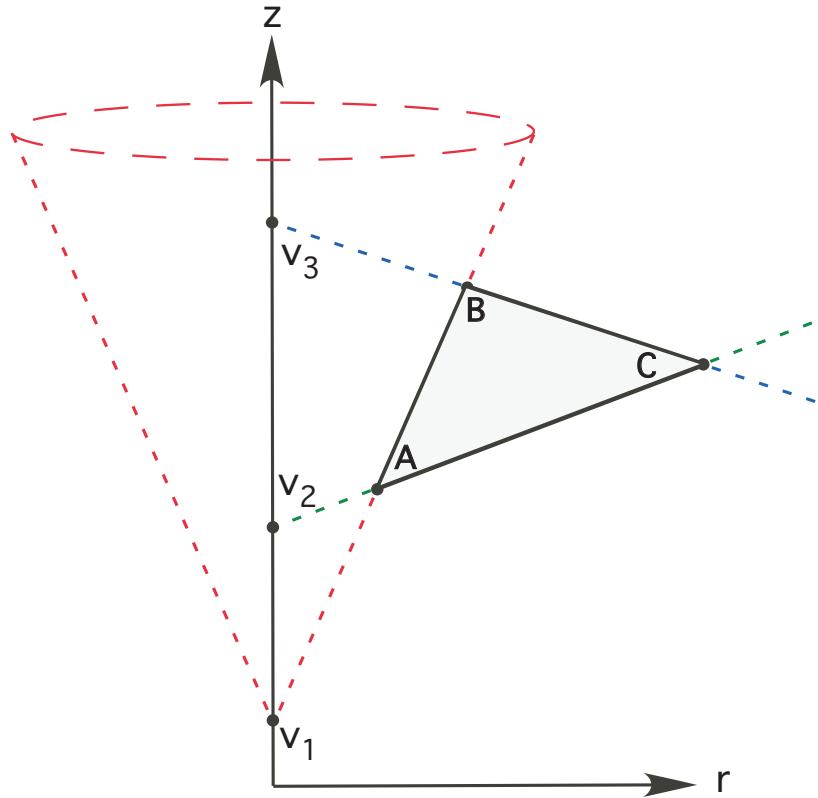}
\par\end{centering}

\caption{In the r-z plane the nodes A, B and C are connected by the Delaunay tessellation to form a triangular cell $\Delta ABC$.  The three extended lines that form the cell edges intercept the z-axis at points $v_1$,$v_2$ and $v_3$.  Upon rotation about the z-axis these lines become the surfaces of three cones with vertices at $v_1$, $v_2$ and $v_3$.  The cone with vertex $v_1$ is shown by the dashed lines.  The interstice formed by the three intersecting cones is a toroid with triangular cross-section.  \label{fig:cell}}

\end{figure}

Figure \ref{fig:cell} shows how the triangular torus is formed from the interstice of three intersecting cones.  Computationally, the most efficient way to transport photons through these volumes is to jump from one cell face (the entry face) to another (the exit face).  In order to determine these locations we need to calculate the intersections between a straight line (photon trajectory) and the three cones that form the surfaces of our toroidal cells (in 3D).   The exit face is always the face with the intersection point that is closest to the current position of the photon (in the direction of photon propagation).

The surface of a cone satisfies the equation,

\begin{equation}
\mathbf{\hat z} \cdot (\mathbf{x}-\mathbf{v})=|\mathbf{x}-\mathbf{v}|\cos\theta,\label{eq:cone}
\end{equation}
which simply states that the vector between a location in space $\mathbf{x}$ and the cone vertex $\mathbf{v}$, makes an angle $\theta$ with the cone axis (in our case the z-axis, with unit direction vector $\mathbf{\hat z}$). We refer to $\theta$ as the opening angle.  It is convenient to square Eqn. \ref{eq:cone} to obtain a quadratic version of the cone equation. There will now be two solutions, one of which corresponds to the reflection cone.  For example, if $\mathbf{x}$ lies on the cone, then so too does the point $2\mathbf{v}-\mathbf{x}$ that simply corresponds to the reflection of $\mathbf{x}$ through the vertex $\mathbf{v}$.  This formulation automatically deals with cones that are `upside down' (i.e. have opening angle $\theta>90\deg$).  Thus, the formulation is comprehensive and it only remains to select the solution corresponding to the appropriate cone orientation.  We proceed by writing the squared Eqn \ref{eq:cone} in matrix form,

\begin{equation}
(\mathbf{x}-\mathbf{v})^TM(\mathbf{x}-\mathbf{v})=0.\label{eq:matrix}
\end{equation}
The matrix $M=(\mathbf{\hat z}\mathbf{\hat z}^T-\cos\theta^2I)$ where $I$ is the unit matrix.  The straight line trajectory of a photon can be expressed parametrically as,

\begin{equation}
\mathbf{x}(t)=\mathbf{p}+t\mathbf{k},\label{eq:line}
\end{equation}
where $\mathbf{k}=(k_x,k_y,k_z)$ is the unit direction vector, $\mathbf{p}$ is a point of origin (the current photon location), and $t$ the distance along the trajectory.   The intersection between the photon trajectory and cell wall is found by inserting Eqn \ref{eq:line} into Eqn \ref{eq:matrix} and solving for the distance $t$.  Upon doing this we obtain the quadratic equation,

\begin{equation}
c_2t^2+c_1t+c_0=0,\label{eq:quad}
\end{equation}
where $c_2=\mathbf{k}^TM\mathbf{k}$,  $c_1=\mathbf{k}^TM(\mathbf{p}-\mathbf{v})$, and $ c_0=(\mathbf{p}-\mathbf{v})^TM(\mathbf{p}-\mathbf{v})$.  The most case is that $c_2\ne0$, making Eqn \ref{eq:quad} a quadratic equation with solutions $t=(-c_1\pm2\sqrt{c_1^2-c_2c_0})/2c_2$.   If $c_1^2-c_2c_0<0$ no real solutions are possible and no intersection exists.  If $c_1^2-c_2c_0=0$ a single (repeated root) intercept is found, which is seen to be tangent to the cone at the point $t=-c_1/c_2$.  If $c_1^2-c_2c_0>0$ two intercepts exist at $-c_1\pm \sqrt{c_1^2-c_2c_0}/c_2$.   There are additional possible situations, for example, $c_2=0$ renders Eqn \ref{eq:quad} linear, implying that only one intersection exists.  In this rather improbable case the photon trajectory is parallel (but not coincident) with a straight line on the cone.  Finally, if $c_2=c_1=c_0=0$ the photon path lies on the cone surface, which is exceedingly unlikely.   A computational implementation should takes these cases into account.

There are extremal cones that may arise from the tessellation and need to be taken into consideration.  If the cell edge in the $r-z$ plane is perpendicular to the z-axis,  $\mathbf{\hat z} \cdot (\mathbf{A}-\mathbf{B}) =0$, then upon rotation we have a horizontal plane rather than a cone.  This can be seen by putting $\theta=90$ into Eqn \ref{eq:cone}.  Finding the intersection of a line and a plane is straightforward.  On the other hand, if the cell edge in the $r-z$ plane is parallel to the z-axis, for example $\mathbf{\hat z} \cdot (\mathbf{A}-\mathbf{B}) =|\mathbf{A}-\mathbf{B}|$,  upon rotation we have a cylinder instead of a cone.





\begin{figure}
\begin{centering}
\includegraphics[scale=0.75]{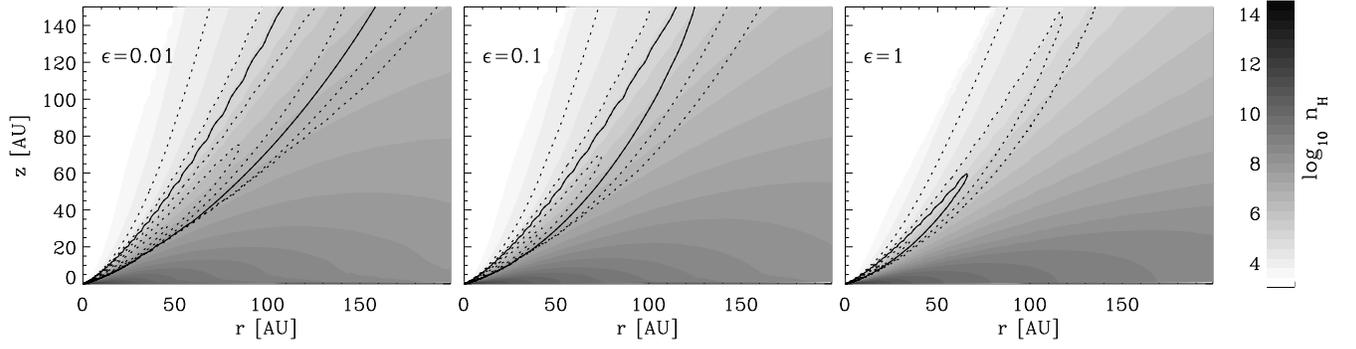}
\par\end{centering}

\caption{Photodissociated layer of atomic H in the three modified D'Alessio disk models.  The filled contours represent the total density, $n_H$. The atomic H density $n(\textrm{H})$ is shown by the unfilled contours, each contour being separated by a factor of $10^{0.5}$. The solid contour corresponds to $n(\textrm{H})=10^5$cm$^3$.\label{fig:columns}}

\end{figure}

\begin{figure}
\begin{centering}
\includegraphics[scale=0.75]{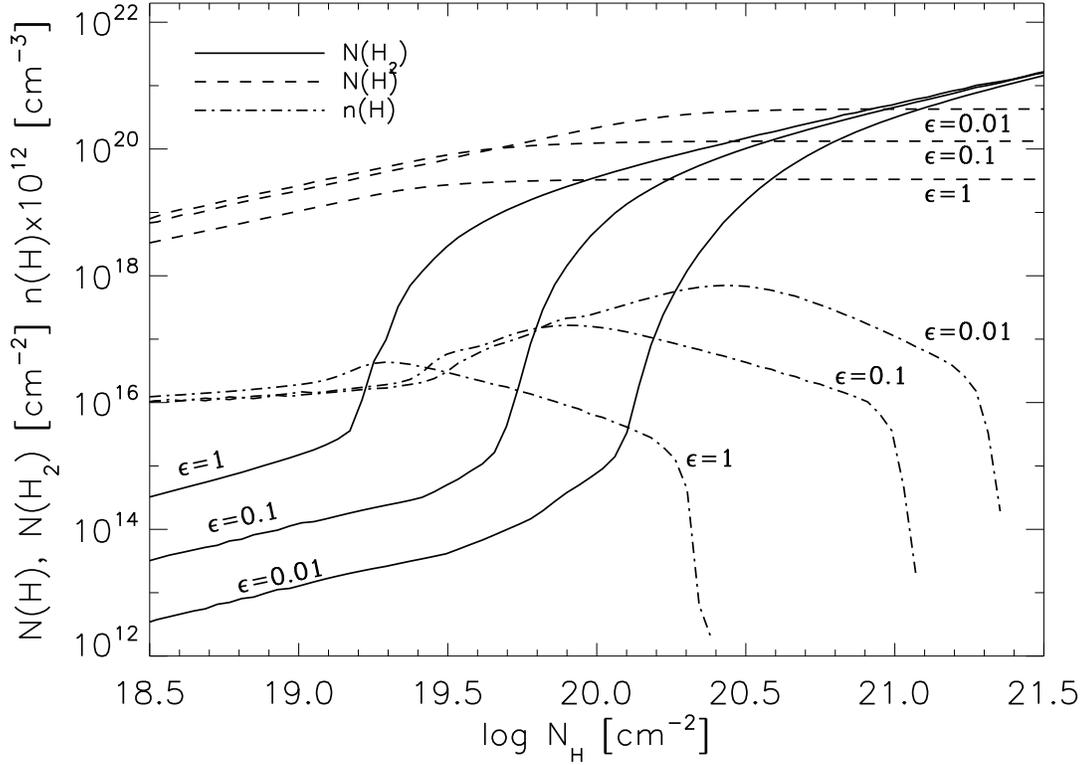}
\par\end{centering}

\caption{The vertical distribution of atomic and molecular hydrogen at $r=100$AU
in the modified D'Alessio disks with varying degrees of dust settling,
$\epsilon$. Vertical column densities of atomic hydrogen ($\colh$,
\textit{dashed line}) and molecular hydrogen ($\colhh$, \textit{solid
line}), and (scaled) volumetric atomic hydrogen density ($\nh$, \textit{dot-dashed
line}) are shown against the total hydrogen column, $\hcol$. Vertical
column densities are integrated downwards from infinity.\label{fig:columns}}

\end{figure}

\end{document}